\documentclass{article}
\pdfoutput=1

\newcommand{\R}{\mathbb{R}}

\newcommand{\mathscr}{\ensuremath\EuScript}
\newcommand{\too}{\ensuremath\mathop{-\!\!\!\longrightarrow}}

\def\hthickline{%
\noalign{\hrule height0.8pt}\arrayrulewidth0pt}

\usepackage{graphicx,amsmath,amssymb,euscript}
\usepackage{natbib}

\author{Graeme K. Ambler\\University of Bristol \and Bernard W. Silverman\\University of Oxford}
\date{2004}

\title{Perfect simulation for Bayesian wavelet thresholding with correlated coefficients}

\begin{document}

\maketitle

\begin{abstract}
We introduce a new method of Bayesian wavelet shrinkage for
reconstructing a signal when we observe a noisy version.  Rather than
making the usual assumption that the wavelet coefficients of the signal are
independent, we allow for the possibility that they are locally correlated
in both location (time) and scale (frequency).  This leads us to a prior 
structure which is, unfortunately, analytically intractable.  Nevertheless, 
it is possible to draw independent samples from a close approximation to the posterior 
distribution by an approach based on
Coupling From The Past, making it possible to use a 
simulation-based approach to fit the model.
\end{abstract}

\section{Introduction}

Consider the the standard nonparametric regression problem
\begin{equation}\label{regr}
	y_i = g(t_i)+\varepsilon_i.
\end{equation}
where we observe a noisy version of an unknown function $g$ at regularly
spaced intervals $t_i$.  The noise, $\varepsilon_i$ is assumed to be 
independent and Normally distributed with zero mean and variance $\sigma^2$.

The standard wavelet-based approach to this problem is based on two 
properties of the wavelet transform:
\begin{enumerate}
\item A large class of ``well-behaved'' functions can be sparsely represented 
in wavelet-space.
\item The wavelet transform transforms independent, identically distributed
noise to independent, identically distributed wavelet coefficients.
\end{enumerate}

These two properties combine to suggest that a good way to remove noise
from a signal is to transform the signal into wavelet space, discard all 
of the small coefficients (i.e. threshold), and perform the inverse 
transform.  Since the true (noiseless) signal had a sparse representation 
in wavelet space, the signal will essentially be concentrated in a small 
number of large coefficients.  The noise, on the other hand, will still be
spread evenly among the coefficients, so by discarding the small 
coefficients we must have discarded mostly noise and will thus have found a 
better estimate of the true signal.

The problem then arises of what to choose as a threshold value.  General 
methods that have been applied in the wavelet context are SureShrink 
\citep{don-joh:ada-unk}, cross-validation \citep[see][]{nas:wav-shr} and 
False discovery rates \citep[see][]{abr-ben:ada-thr}.  The BayesThresh 
approach \citep{abr-sap-sil:wav-thr} proposes a Bayesian hierarchical
model for the wavelet coefficients, using a mixture of a point mass at $0$ 
and a $N(0,\tau^2)$ density as their prior.  The marginal posterior median 
of the population wavelet coefficient is then used as the estimate.  This 
gives a thresholding rule, since the point mass at $0$ in the prior gives 
non-zero probability that the population wavelet coefficient will be zero.

Most Bayesian approaches to wavelet thresholding model the coefficients 
independently.  In order to capture the notion that nonzero wavelet coefficients may 
be in some way clustered, we allow prior dependency between the coefficients by
modelling them using an extension of the area-interaction process of 
\citet{bad-lie:are-int}.  The basic idea is that if a coefficient is nonzero then 
it is more likely that its neighbours (in a suitable sense) are also non-zero.

The disadvantage of this prior is that it is no longer possible to compute the 
estimates explicitly, and a method like Markov chain Monte Carlo (MCMC) has to be 
used.  A key problem with the use of MCMC is that one can rarely be absolutely sure 
that the Markov chain which is used for a given simulation has converged to its
stationary distribution.  \citet{pro-wil:exa} introduced coupling from the past 
(CFTP) as an approach to solve this problem and produce exact realisations from the 
stationary distribution of a Markov chain.  We use an extension of this method to 
sample from the posterior distribution of our model.

An outline of the paper is as follows.  In Section~\ref{model} we briefly survey 
the area-interaction process and introduce our model for the wavelet coefficients.  
In Section~\ref{simulation} we discuss coupling from the past, and an extension 
which allows us to sample from the posterior distribution of our model.  In 
Section~\ref{simstud} we present a simulation study to compare our method with the
others introduced in this section.  Section~\ref{concl} presents some conclusions 
and discusses possible avenues for future work.

\section{A Bayesian model for wavelet thresholding}\label{model}

\subsection{The Area-interaction point process}\label{are-int}

The area-interaction point process \citep{bad-lie:are-int} is a spatial point
process capable of producing both moderately clustered and moderately ordered
patterns dependent on the value of its clustering parameter.  It was
introduced primarily to fill a gap left by the Strauss point process
\citep{str:a-mod}, which can only produce ordered point patterns
\citep{kel-rip:a-not}.

The definition of the area-interaction process depends on a specification of
the \emph{neighbourhood} of any point in the space $\chi$ on which the process
is defined.  Given any $x\in\chi$ we denote by $B(x)$ the neighbourhood of the
point $x$.  Given a set $X\subseteq\chi$, the neighbourhood $U(X)$ of $X$ is
defined as $\bigcup_{x\in X}B(x)$.  The general area-interaction process
is defined by \citeauthor{bad-lie:are-int} as follows.

Let $\chi$ be some locally compact complete metric space and
$\mathfrak{R}^f$ be the space of all possible configurations of points in
$\chi$.  Suppose that $m$ be a finite Borel regular measure on $\chi$ and
$B:\chi\to\mathscr{K}$ be a myopically continuous function 
\citep{mat:ran}, where $\mathscr{K}$ is the class of all compact subsets 
of $\chi$.  Then the probability density of the general area-interaction 
process is given by
\begin{equation}\label{gaip}
	p(X) = \alpha\lambda^{N(X)}\gamma^{-m\{U(X)\}}
\end{equation}
with respect to the unit rate Poisson process, where $N(X)$ is the number of
points in configuration $X=\{x_1,\ldots,x_{N(X)}\}\in\mathfrak{R}^f$, $\alpha$
is a normalising constant and $U(X) = \bigcup_{i=1}^{N(X)}B(x_i)$ as above.
  
In the following section we define the particular special case of this point 
process that we use as our prior.  In the context of the rest of the paper, 
$\chi$ is a discrete space, so the technical conditions required of 
$m(\cdot)$ and $B(\cdot)$ are trivially satisfied.

\subsection{Model specification}\label{ext}


\citet{abr-sap-sil:wav-thr} consider the problem where the true
wavelet coefficients are observed subject to Gaussian noise with zero mean and some
variance $\sigma^2$,
\[
	\widehat{d}_{jk}|d_{jk} \sim N(d_{jk},\sigma^2),
\]
where $\widehat{d}_{jk}$ is the value of the noisy wavelet coefficient
(the data) and $d_{jk}$ is the value of the true coefficient.

Their prior distribution on
the true wavelet coefficients is a mixture of a Normal distribution with
zero mean and variance dependent on the level of the
coefficient, and a point mass at zero as follows:
\begin{equation}\label{BayesThresh}
	d_{jk}\sim\pi_{j} N(0,\tau_j^2)+(1-\pi_{j})\delta(0),
\end{equation}
where $d_{jk}$ is the value of the $k$th coefficient at level $j$ of the
discrete wavelet transform, and the mixture weights $\{\pi_j\}$ are
constant within each level.  An alternative formulation of this can be
obtained by introducing auxiliary variables $Z=\{\zeta_{jk}\}$ with
$\zeta_{jk}\in\{0,1\}$ and independent hyperpriors
\begin{equation}\label{indep}
	\zeta_{jk}\sim \text{Bernoulli}(\pi_j).
\end{equation}
The prior given in equation (\ref{BayesThresh}) is then expressed as
\begin{equation}\label{SpaBayesThresh}
	d_{jk}|Z\sim N(0,\zeta_{jk}\tau_j^2).
\end{equation}

The starting point for our extension of this approach is to note that $Z$ can be considered
as being a point process on the discrete space, or lattice, $\chi$ of indices $(j,k)$ of the wavelet coefficients.
The points of $Z$ give the locations at which the prior variance of the wavelet coefficient, conditional on $Z$,
is nonzero.
From this point of view, the hyperprior structure given in equation~(\ref{indep}) is 
equivalent to specifying $Z$ to be a Binomial process with rate function $p(j,k)=\pi_j$.  

Our general approach will be to replace $Z$ by a more general lattice process 
$\xi$ on $\chi$.
We allow $\xi$ to have multiple points at particular locations $(j,k)$, so that the number $\xi_{jk}$ 
of points at $(j,k)$ will be a non-negative integer, not necessarily confined to $\{ 0, 1 \}$. 
We will assume that the prior variance is 
proportional to the number of points of $\xi$ falling at the corresponding 
lattice location.   So if there are no points, the prior will be 
concentrated at zero and the corresponding observed wavelet will be 
treated as pure noise;  on the other hand, the larger the number of 
points, the larger the prior variance and the less shrinkage applied to the 
observed coefficient.
To allow for this generalisation, we extend (\ref{SpaBayesThresh}) in the natural way to
\begin{equation}\label{truecoef}
	d_{jk}| \xi \sim N(0,\tau^2 \xi_{jk}),
\end{equation}
where $\tau^2$ is a constant.  

\begin{figure}
	\begin{center}
		\resizebox{0.95\textwidth}{!}{\includegraphics{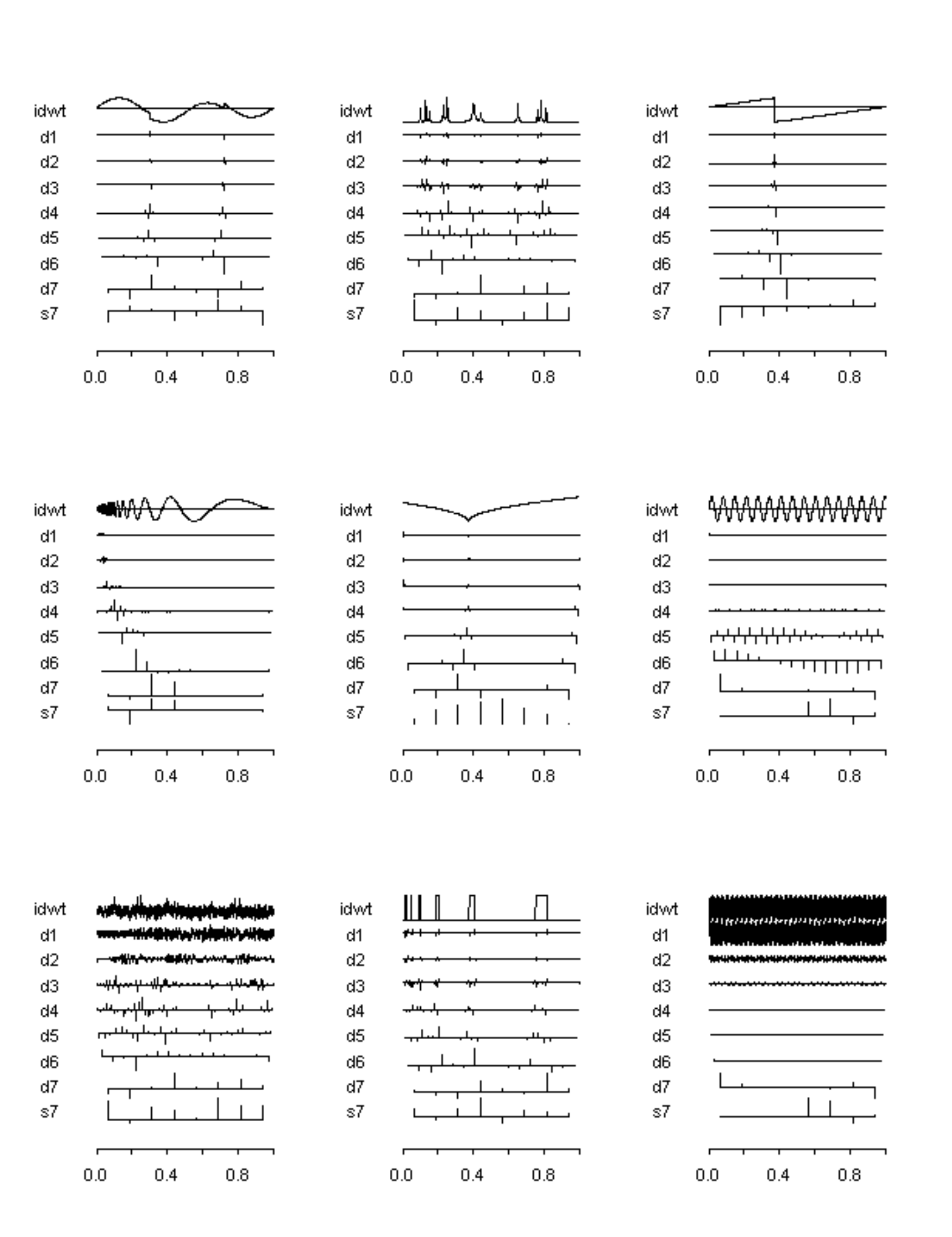}}
	\end{center}
\caption[Examples of the DWT of some test functions.]{Examples of the discrete wavelet transform of some test functions.  There is clear evidence of clustering in most of the graphs.  The original functions are shown above their discrete wavelet transform each time.\label{fig:dwt}}
\end{figure}

We now consider the specification of the process $\xi$.  
While it is natural to expect that the wavelet transform will produce a sparse
representation, the time-frequency localisation properties of the transform
also make it natural to expect that the representation will be clustered in
some sense.  The existence of this clustered structure can be seen clearly in
Figure~\ref{fig:dwt}, which shows the discrete wavelet transform of several
common test functions represented in the natural binary tree configuration.
With this clustering in mind, we model $\xi$ as an area-interaction process on the space $\chi$.  
The choice of the 
neighbourhoods $B(x)$ for $x$ in $\chi$ will be discussed below.   Given the choice of neighbourhoods, the process will 
be defined by 
\begin{equation}\label{aipprior}
	p(\xi) = \alpha\lambda^{N(\xi)}
				\gamma^{-m\{U(\xi)\}}
\end{equation}
where $p(\xi)$ is the intensity relative to 
the unit rate independent auto-Poisson process
\citep{cre:sta}.  If
we take $\gamma>1$ this gives a clustered configuration.  Thus we would expect
to see clusters of large values of $d_{jk}$ if this were a reasonable model
--- which is exactly what we do see in Figure~\ref{fig:dwt}.

A simple application of Bayes theorem tells us that the posterior for our
model is
\begin{eqnarray}\nonumber
	p(\xi ,\mathbf{d}|\widehat{\mathbf{d}}) & = &
		p(\xi)
		\prod_{j,k}p(d_{jk}|\xi_{jk})
		\prod_{j,k}p(\widehat{d}_{jk}|d_{jk},\xi_{jk})\\
	& = &	\alpha\lambda^{N(\xi)}\gamma^{-m\{U(\xi)\}}
		\prod_{j,k}\frac{\exp(-d_{jk}^2/2\tau^2 \xi_{jk})}
			{\sqrt{2\pi\tau^2 \xi_{jk}}}
		\prod_{j,k}\frac{\exp\left\{-(\widehat{d}_{jk}-d_{jk})^2/2\sigma^2\right\}}
			{\sqrt{2\pi\sigma^2}}.\label{post}
\end{eqnarray}

Clearly (\ref{post}) is not a standard density, and in Section \ref{simulation} we will 
discuss an extension 
of coupling from the past which will help us to sample from it.  

\subsection{Specifying the neighbourhood structure}

\begin{figure}
	\begin{center}
		\resizebox{0.95\textwidth}{!}{\includegraphics{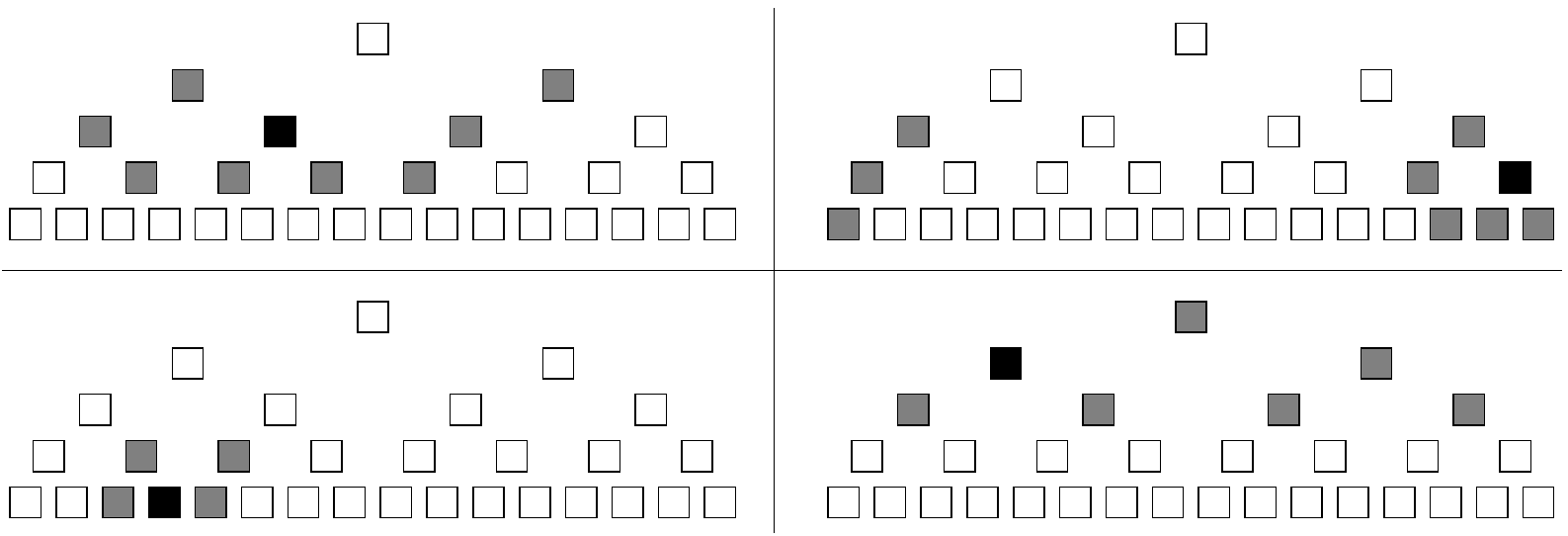}}
	\end{center}
	\caption[Examples of $B(\cdot)$.]{The four plots give examples of what we used as $B(\cdot)$ for four different example locations showing how we dealt with boundaries.  Grey boxes are $B(x)\setminus\{x\}$ for each example location $x$, while $x$ itself is shown as black.\label{fig:tree}}
\end{figure}

In order to complete the specification of our area-interaction prior for $\xi$, 
we need a suitable interpretation of the neighbourhood of a location 
$x= (j,k)$ on the lattice $\chi$ of indices $(j,k)$ of wavelet coefficients.
This lattice is a binary tree, and there are 
many possibilities.  We decided to use the parent, the
coefficient on the parent's level of the transform which is next-nearest
to $x$, the two adjacent coefficients on the level of $x$, the two
children and the coefficients adjacent to them, making a total of nine
coefficients (including $x$ itself).  Figure \ref{fig:tree} illustrates
this scheme, which captures the localisation of both time and frequency
effects.  Figure \ref{fig:tree} also shows
how we dealt with boundaries: we assume that the signal we are
examining is periodic, making it natural to have periodic boundary
conditions in time.
If $B(x)$ overlaps with a frequency boundary we simply discard those parts
which have no locations associated with them.  The simple counting measure
used has $m\{B(x)\}=9$ unless $x$ is in the bottom row or one of the top two
rows.

Other possible neighbourhood functions include using only the parent, children
and immediate sibling and cousin of a coefficient as $B(x)$, or a variation on
this taking into account the length of support of the wavelet used.  Though
we have chosen to use periodic boundary conditions, our method is equally
applicable without this assumption.

\section{Simulation}\label{simulation}

In this section, we develop a practical approach to simulation from a close approximation to the
posterior density (\ref{post}).   We begin by reviewing the general approach of coupling from the past, and then
explore the way in which this concept can be applied to our particular application.

\subsection{Coupling from the past}\label{CFTP}


The motivation behind coupling from the past (CFTP) is the following.  
Suppose that it is desirable to sample from the stationary distribution of an
ergodic Markov chain $\{Z_t\}$ on some (finite) state space $X$ with states 
$1,\ldots,n$.  It is clear that if it were possible to go back an infinite 
amount in time, start the chain running (in state $Z_{-\infty}$) and then 
return to the present, the chain would (with probability 1) be in its 
stationary distribution when one returned to the present (i.e. $Z_0\sim\pi$, 
where $\pi$ is the stationary distribution of the chain).  This is clearly 
not feasible in practice!  \citet{pro-wil:exa} showed that in fact we can 
achieve the same goal by going back a finite (random) amount of time only.

Consider a finite state space with $n$ states, and that we set not one, but
$n$ chains $\{Z^{(1)}_t\}, \ldots, \{Z^{(n)}_t\}$ running at a fixed time 
$-M$ in the past, where $Z^{(i)}_{-M}=i$ for each chain $\{Z^{(i)}_t\}$.  Now 
let all the chains be coupled \citep[see][]{lin:lec} so that if 
$Z^{(i)}_{s}=Z^{(j)}_{s}$ at any time $s$ then 
$Z^{(i)}_{t}=Z^{(j)}_{t} \hspace{1em} \forall t\geq s$.  Then if 
all the chains ended up in the same state $j$ at time zero 
(i.e. $Z^{(i)}_0=j \hspace{1em} \forall i\in X$), we would know that 
whichever state a chain passing from time minus infinity to zero was in 
at time $-M$, the chain would end up in state $j$ at time zero.  Thus $j$ 
must be a sample from the stationary distribution of the Markov chain in 
question.

\citet{ken:are} extended CFTP to cover simulation of the area-interaction 
process discussed in Section~\ref{are-int}, which has an infinite state 
space.  The method makes use of a monotone coupling and stochastic 
domination.  A coupling is monotone if whenever $Z^{(i)}_t\geq Z^{(j)}_t$ 
then $Z^{(i)}_{t+k}\geq Z^{(j)}_{t+k}\;\;\forall k>0$.  Given a monotone
coupling and unique minimum and maximum elements we need only simulate 
Markov chains starting in the maximum and minimum states and check that these 
two have coalesced at time $0$, since chains starting in all other states 
would be sandwiched between these two.  As there is no natural maximum 
element for the area-interaction process, \citeauthor{ken:are} used a Poisson 
process which stochastically dominates the area interaction process of 
interest to generate a maximum process.  More recently, \citet{ken-mol:per}
extended these techniques to more general classes of point processes.  

\citet{amb-sil:per} explain why the method of \citet{ken-mol:per} is not practically feasible for
some classes of point process models, and
provide an alternative method which makes it possible to 
simulate from densities such as equation~(\ref{post}).
We describe their method in the following section.

\subsection{Perfect simulation of spatial point processes}\label{per-sim}

We now move to spatial point processes defined on a set like the 
unit square $[0,1]\times[0,1]\subseteq\R^2$.  Suppose that we wish to sample 
from such a spatial point process with density
\[
	p(X) = \alpha\prod_{i=1}^m f_i(X),
\]
with respect to the unit rate Poisson process, where $\alpha\in(0,\infty)$ and 
$f_i:\mathfrak{R}^f\to\R$ are positive valued functions which are (a) monotonic 
with respect to the partial ordering induced by the subset relation, i.e. for 
any $X$ and $Y$ related by $X\subseteq Y$, $f_i(X)\leq f_i(Y)\;\forall i$, and 
(b) whose conditional intensity
\[
	\lambda_f(u;X) = \frac{f(X\cup\{u\})}{f(X)}
\]
is uniformly bounded.  \citet{amb-sil:per} show that this is possible 
using the following algorithm.

Let $D$ be a two-dimensional Poisson process with rate equal to 
\begin{equation}\label{gen_dom}
\lambda = \prod_{i=1}^m\max_{X,\{x\}}\lambda_{f_i}(x;X),
\end{equation}
evolving over time according to a birth-death process with birth rate equal 
to (\ref{gen_dom}) and unit death rate.  Let $D(T)$ be the configuration of
points in process $D$ at time $T$.  For simplicity of notation, constrain 
this function to be right-continuous, so that if there is a birth in $D$ at 
time $T$ then $D(T)$ is the configuration which existed in $D$ immediately 
prior to the birth.

Now let $U$ be a birth-death process which is started from an initial 
configuration equal to that of $D$ at some time in the past, and $L$ be a 
birth-death process which is started from an initial configuration equal to 
a thinned version of $D$, where points are accepted with probability
\[
	\frac1\lambda\prod_{i=1}^m\min_{X,\{x\}}\lambda_{f_i}(x;X)
\]

The processes $U$ and $L$ evolve through time as follows.  If a point 
$\{u\}$ is born in $D$ at time $T$ then $\{u\}$ is also born in $U$ at time 
$T$ with probability
\begin{equation}\label{gen_up}
	\frac1\lambda\prod_{i=1}^m \max\left\{ \lambda_{f_i}[u;U(T)] , 
			\lambda_{f_i}[u;L(T)]\right\}.
\end{equation}
The point $\{u\}$ is born in $L$ at time time $T$ with probability
\begin{equation}\label{gen_low}
	\frac1\lambda\prod_{i=1}^m \min\left\{ \lambda_{f_i}[u;U(T)] , 
			\lambda_{f_i}[u;L(T)]\right\}.
\end{equation}
If a point dies in $D$ then if it existed in $U$ or $L$ then it dies there 
also.

Finally, generate $D$ backwards in time from zero to some time $-T$ and start 
$U$ and $L$ there.  Now run them forward to time zero.  If $U(0)=L(0)$ then 
the configuration $U(0)$ (or equivalently $L(0)$) is a sample from the 
required spatial point process.  If not, we must generate $D$ further back in 
time and try again, keeping the probabilities used for acceptance/rejection 
in the first round.

In our case, we are simulating from a process on a lattice rather than the unit square, and
in the next section, we set out a modified version more appropriate to that context.

\subsection{Exact posterior sampling for lattice processes}\label{alg}

One of the advantages of the Normal model we propose in Section~\ref{ext} 
is that it is possible to integrate out $d_{jk}$ and work only with the lattice process
$\xi$.  Performing this calculation, we see that equation~(\ref{post}) 
can be rewritten as
\begin{eqnarray*}
	p(\xi|\widehat{\mathbf{d}}) &\!\!\!=\!\!\!& p(\xi)
		\prod_{j,k}\int\frac{\exp(-d_{jk}^2/2\tau^2\xi_{jk})}
			{\sqrt{2\pi\tau^2\xi_{jk}}}
			\frac{\exp\left\{-(\widehat{d}_{jk}-d_{jk})^2/2\sigma^2\right\}}
			{\sqrt{2\pi\sigma^2}}dd_{jk}\\
	&\!\!\!=\!\!\!&	p(\xi)\prod_{j,k}\int
			\frac{\exp\left[-\left\{d_{jk}^2(\sigma^2+\tau^2\xi_{jk})
				-2d_{jk}\widehat{d_{jk}}\tau^2\xi_{jk}
				+\widehat{d_{jk}}\tau^2\xi_{jk}\right\}/2\tau^2\xi_{jk}\sigma^2\right]}
			{\sqrt{4\pi^2\tau^2\xi_{jk}\sigma^2}}dd_{jk}\\
	&\!\!\!=\!\!\!&	p(\xi)\prod_{j,k}
		\frac{\exp\left\{\frac{-\widehat{d_{jk}}^2}{2(\sigma^2+\tau^2\xi_{jk})}\right\}}
		{\sqrt{4\pi^2\tau^2\xi_{jk}\sigma^2}}
			\int\exp\left\{
			-\left(\frac{\sigma^2+\tau^2\xi_{jk}}{2\tau^2\xi_{jk}\sigma^2}\right)
			\left(d_{jk}-\frac{\widehat{d_{jk}}\tau^2\xi_{jk}}{\sigma^2+\tau^2\xi_{jk}}
				\right)^2\right\}dd_{jk}\\
	&\!\!\!=\!\!\!&	p(\xi)\prod_{j,k}
		\frac{\exp\left\{\frac{-\widehat{d_{jk}}^2}{2(\sigma^2+\tau^2\xi_{jk})}\right\}}
		{\sqrt{4\pi^2\tau^2\xi_{jk}\sigma^2}}
		\left(\frac{2\pi\sigma^2\tau^2\xi_{jk}}{\sigma^2+\tau^2\xi_{jk}}\right)^{1/2}\\
	&\!\!\!=\!\!\!&	p(\xi)\prod_{j,k}\frac{\exp
			\left\{-\widehat{d_{jk}}^2/2(\sigma^2+\tau^2\xi_{jk})\right\}}
		{\sqrt{2\pi(\sigma^2+\tau^2\xi_{jk})}}.
\end{eqnarray*}
We now see that it is possible to sample from the posterior by simulating only 
the process $\xi$ and ignoring the marks $\mathbf{d}$.  This lattice 
process is amenable to perfect simulation using the method of 
\citet{amb-sil:per}.  Let
\begin{align*}
	f_1(\xi)= &\; \lambda^{N(\xi)},\\
	f_2(\xi)= &\; \gamma^{-m\left\{U(\xi)\right\}},\\
	f_3(\xi)= &\; \prod_{j,k}
		\exp\left\{-\widehat{d_{jk}}^2\left/\right.2(\sigma^2+\tau^2\xi_{jk})\right\}
		\text{ and}\\
	f_4(\xi)= &\; \prod_{j,k}\left\{2\pi(\sigma^2+\tau^2\xi_{jk})\right\}^{-1/2}.
\end{align*}
Then
\begin{align*}
	\lambda_{f_1}(u;\xi)= &\; \lambda,\\
	\lambda_{f_2}(u;\xi)= &\; \gamma^{-m\left\{B(u)\setminus U(\xi)\right\}}\leq1,\\
	\lambda_{f_3}(u;\xi)= &\; 
		\exp\left[\frac{\widehat{d_{u}}^2\tau^2}
			{2(\sigma^2+\tau^2\xi_{u})\left\{\sigma^2+\tau^2(\xi_{u}+1)\right\}}\right]
			\leq\exp\left\{\frac{\widehat{d}_u^2\tau^2}
			{2\sigma^2(\tau^2+\sigma^2)}\right\}
\text{ and}\\
	\lambda_{f_4}(u;\xi)= &\;
		\left\{\frac{\tau^2\xi_u+\sigma^2}{\tau^2(\xi_u+1)+\sigma^2}\right\}^{1/2}
			\leq1.
\end{align*}
By a slight abuse of notation, in the second and third equations above we use 
$u$ to refer both to the point $\{u\}$ and the location $(j,k)$ at which it 
is found.
The functions $f_1,\ldots,f_4$ are also monotone with respect to the subset relation, 
so all of the conditions for exact simulation using the method of
\citet{amb-sil:per} are satisfied.

In the spatial processes considered in detail in \citet{amb-sil:per}, the 
dominating process had constant intensity across the space $\chi$. 
In the present context, however, it is necessary in practice to  use a dominating process which 
has a different rate at each lattice location, and then use 
location-specific maxima
and minima rather than global maxima and minima.   Because we can now 
use location-specific, rather than global,
maxima and minima, we can initialise upper and lower processes that are much
closer together than would have been possible with a constant-rate 
dominating process, and consequently reducing coalescence times
to feasible levels.
A constant rate dominating process would not have been 
feasible due to the size of
the global maxima, so this modification to the method of \citet{amb-sil:per} 
is essential; see Section~\ref{large+small1} for details.  
\citet[Chapter 5]{amb:dom-cou} gives some other 
examples of dominating processes with location-specific intensities.


The location-specific rate of the dominating process $D$ is 
\begin{equation}\label{lambda_dom}
	\lambda_{jk}^{dom} = \lambda 
		       e^{\widehat{d}_{jk}^2\tau^2/2\sigma^2(\tau^2+\sigma^2)}
\end{equation}
for each location $(j,k)$ on the lattice.  The lower process is then started 
as a thinned version of $D$.  Points are accepted with probability
\[
	P(x) = \gamma^{-M(\chi)}\left(\frac{\sigma^2}{\tau^2+\sigma^2}\right)^{1/2}
		\times\exp\left\{ - 
			\frac{\widehat{d}_x^2\tau^2}
				{2\sigma^2(\tau^2+\sigma^2)}\right\},
\]
where $M(\chi) = \max_\chi[m\{B(x)\}]$.
The upper and lower processes are then evolved through time, 
accepting points as described in Section~\ref{per-sim} with probability
\[
\frac1{\lambda_{jk}^{dom}}\lambda_{f_1}(u;\xi^{\text{up}})
		\lambda_{f_2}(u;\xi^{\text{up}})
		\lambda_{f_3}(u;\xi^{\text{low}})
		\lambda_{f_4}(u;\xi^{\text{up}})
\]
for the upper process and
\[
\frac1{\lambda_{jk}^{dom}}\lambda_{f_1}(u;\xi^{\text{low}})
		\lambda_{f_2}(u;\xi^{\text{low}})
		\lambda_{f_3}(u;\xi^{\text{up}})
		\lambda_{f_4}(u;\xi^{\text{low}})
\]
for the lower process.  The remainder of the algorithm carries
over in the obvious way.  
There are still some issues to be addressed due to very high birth rates 
in the dominating process, and this will be done in Section \ref{large+small1}.

\subsection{Using the generated samples}\label{using}

Although $\mathbf{d}$ was integrated out for simulation reasons in
Section~\ref{ext} it is, naturally, the quantity of interest.  Having
simulated realisations of $\xi|\widehat{\mathbf{d}}$ we then
generate $\mathbf{d}|\xi,\widehat{\mathbf{d}}$ for each
realisation $\xi$ generated in the first step.  The sample
median of $\mathbf{d}|\xi,\widehat{\mathbf{d}}$ gives an estimate for
$\mathbf{d}$.  The median is used instead of the mean as this
gives a thresholding rule \citep[defined by][as a rule giving $p(d_{jk}=0|\widehat{\mathbf{d}})>0$]{abr-sap-sil:wav-thr}.

We calculate $p(\mathbf{d}|\xi,\widehat{\mathbf{d}})$ using logarithms
for ease of notation.  Assuming that $\xi_{jk}\ne 0$ we find
\begin{eqnarray*}
	\log p(d_{jk}|\widehat{d}_{jk},\xi_{jk}\ne 0) & = &
		\log{p(d_{jk}|\xi_{jk}\ne0)} +
		\log{p(\widehat{d}_{jk}|d_{jk},\xi_{jk}\ne0)} + C\\
	& = & \frac{-d_{jk}^2}{2\tau^2\xi_{jk}}+
			\frac{-(\widehat{d}_{jk}-d_{jk})^2}{2\sigma^2} + C_1\\
	& = & -\frac{(\sigma^2+\tau^2\xi_{jk})
		\left(d_{jk}-\frac{\tau^2\xi_{jk} \widehat{d}_{jk}}
			{\sigma^2+\tau^2\xi_{jk}}\right)^2}
		{2\sigma^2\tau^2\xi_{jk}}+C_2
\end{eqnarray*}
where $C$, $C_1$ and $C_2$ are constants.  Thus
\[
	d_{jk}|\widehat{d}_{jk},\xi_{jk}\ne 0 \sim
		N\left(\frac{\tau^2\xi_{jk} \widehat{d}_{jk}}
			{\sigma^2+\tau^2\xi_{jk}},
		\frac{\sigma^2\tau^2\xi_{jk}}{\sigma^2+\tau^2\xi_{jk}}\right).
\]
When $\xi_{jk}=0$ we clearly have $p(d_{jk}|\xi_{jk},\widehat{d}_{jk})=0$.


\subsection{Dealing with large and small rates}\label{large+small1}

We now deal with some approximations which are necessary to allow our algorithm to be feasible computationally.
Recall from equation (\ref{lambda_dom})
that if the maximum data value $d_{jk}$ is twenty times larger in magnitude
than the standard deviation of the noise (a not uncommon event for
reasonable noise levels) then we have
\begin{eqnarray*}
	\lambda_{dom} & = & \lambda e^{400\sigma^2\tau^2/
						2\sigma^2(\tau^2+\sigma^2)}\\
		& = & \lambda e^{200\tau^2/(\tau^2+\sigma^2)}.
\end{eqnarray*}
Now unless $\tau$ is significantly smaller than $\sigma$, this will result in
enormous birth rates, which make it necessary to modify the algorithm appropriately.  
To address this issue, we noted that 
the chances of there being no
live points at a location whose data value is large (resulting in a value
of $\lambda_{dom}$ larger than $e^4$) is sufficiently small that for the
purposes of calculating $\lambda_{f_2}(u;\xi)$ for
nearby locations it can be assumed that the number of points alive was
strictly positive.  

This means that we do not know the true value
of $\xi_{jk}$ for the locations with the largest values of $d_{jk}$.  This
leads to problems since we need to generate $d_{jk}$ from the distribution
\[
	d_{jk}|\xi_{jk},\widehat{d}_{jk}
		\sim
			N\left(\frac{\tau^2\xi_{jk}\widehat{d}_{jk}}
				{\sigma^2+\tau^2\xi_{jk}},
			\frac{\sigma^2\tau^2\xi_{jk}}
				{\sigma^2+\tau^2\xi_{jk}}\right),
\]
which requires values of $\xi_{jk}$ for each location $(j,k)$ in the
configuration.
To deal with this issue, we first note that
\[
	\frac{\tau^2\xi_{jk}\widehat{d}_{jk}}{\sigma^2+\tau^2\xi_{jk}}
		\hspace{.5em} \too_{\xi_{jk}\to\infty} \hspace{.5em}
			\widehat{d}_{jk}
\]
monotonically from below, and that
\[
	\frac{\tau^2\xi_{jk}\sigma^2}{\sigma^2+\tau^2\xi_{jk}}
		\hspace{.5em} \too_{\xi_{jk}\to\infty} \hspace{.5em}
			\sigma^2,
\]
also monotonically from below.  Since $\sigma$ is typically small,
convergence is very fast indeed.  Taking $\tau=\sigma$ as an example we see
that even when $\xi_{jk}=5$ we have
\[
	\frac{\tau^2\xi_{jk}\widehat{d}_{jk}}{\sigma^2+\tau^2\xi_{jk}}
		=
			\frac56\widehat{d}_{jk}
\]
and
\[
	\frac{\tau^2\xi_{jk}\sigma^2}{\sigma^2+\tau^2\xi_{jk}}
		=
			\frac56\sigma^2.
\]
We see that we are already within $\frac16$
of the limit.  Convergence is even faster for larger values of $\tau$.

We also recall that the dominating process
gives an upper bound for the value of $\xi_{jk}$ at every location.  Thus a
good estimate for $d_{jk}$ would be gained by taking the value of $\xi_{jk}$
in the dominating process for those points where we do not know the exact
value.  This is a good solution but is unnecessary in some cases, as
sometimes the value of $\lambda_{dom}$ is so large that there is little
advantage in using this value.
Thus for exceptionally large values of $\lambda_{dom}$ we simply use
$N(\widehat{d}_{jk},\sigma^2)$ numbers as our estimate of $d_{jk}$.

\section{Simulation Study}\label{simstud}

We now present a simulation study of the performance of our
estimator relative to several established Wavelet-based estimators.  Similar
to the study of \citet{abr-sap-sil:wav-thr}, we investigate the 
performance of our method on the four standard test functions of 
\citet{don-joh:ide-spa,don-joh:ada-unk}, 
namely ``Blocks'', ``Bumps'', ``Doppler'' and ``Heavisine''.  These test 
functions are used because they exhibit different kinds behaviour typical of 
signals arising in a variety of applications.  

The test functions were simulated at 256 points equally spaced on the unit
interval.  The test signals were centred and scaled so as to have mean value
$0$ and standard deviation $1$.  We then added independent $N(0,\sigma^2)$
noise to each of the functions, where $\sigma$ was taken as $1/10$, $1/7$
and $1/3$.  The noise levels then correspond to root signal-to-noise ratios
(RSNR) of $10$, $7$ and $3$ respectively.  We performed 25 replications.  For
our method, we simulated 25 independent draws from the posterior distribution
of the $d_{jk}$'s and used the sample median as our estimate, as this gives a
thresholding rule.  For each of the runs, $\sigma$ was set to the standard
deviation of the noise we added, $\tau$ was set to $1.0$, $\lambda$ was set
to $0.05$ and $\gamma$ was set to $3.0$.

The values of parameters $\sigma$ and $\tau$ were
set to the true values of the standard deviation of the noise and the
signal, respectively.  In practice it will be necessary to develop some
method for estimating these values.  The value of $\lambda$ was chosen to
be $0.05$ because it was felt that not many of the coefficients would be
significant.  The value of $\gamma$ was chosen based on small trials for the
heavisine and jumpsine datasets.

We compare our method with several established wavelet-based estimators
for reconstructing noisy signals: SureShrink \citep{don-joh:ide-spa},
two-fold cross-validation as applied by \citet{nas:wav-shr}, ordinary
BayesThresh \citep{abr-sap-sil:wav-thr}, and the false
discovery rate as applied by \citet{abr-ben:ada-thr}.

For test signals ``Bumps'', ``Doppler'' and
``Heavisine'' we used Daubechies least asymmetric wavelet of order 10
\citep{dau:ten}.  For ``Blocks'' we used the Haar wavelet, as the
original signal was piecewise constant.  The analysis was carried out
using the freely available $R$ statistical package.  The WaveThresh package
\citep{nas:wav-thr} was used to perform the discrete wavelet transform and
also to compute the SureShrink, cross-validation, BayesThresh and false
discovery rate estimators.

\begin{table}
\caption[Comparison of our estimator with other wavelet-based estimators]{\label{tab:sim}Average mean-square errors ($\times 10^4$) for the area-interaction BayesThresh (AIBT), SureShrink (SS), 
cross-validation (CV), ordinary BayesThresh (BT) and false discovery rate (FDR) estimators for four test functions for three values of the root signal-to-noise ratio.  
Averages are based on 25 replicates.  
Standard errors are given in parentheses.}
\centering
 \begin{tabular}{ccr@{ }rr@{ }rr@{ }rr@{ }r@{\hspace{.7em}}c}
 \hthickline
  RSNR	& Method	& \multicolumn{9}{c}{Test functions}\\ \cline{3-11}
	&	& \multicolumn{2}{c}{Blocks}   & \multicolumn{2}{c}{Bumps}    & \multicolumn{2}{c}{Doppler} & \multicolumn{3}{c}{Heavisine}\\ \hline
 	& AIBT	& 25 & (1)    & 84 & (2)    & 49 & (1)   & \hspace{1em}32 & (1)&\\
  	& SS	& 49 & (2)    & 131 & (6)   & 54 & (2)   & 66 & (2)\\
  10	& CV	& 55 & (2)    & 392 & (21)  & 112 & (5)  & 31 & (1)\\
  	& BT	& 344 & (10)  & 1651 & (17) & 167 & (5)  & 35 & (2)\\
  	& FDR	& 159 & (14)  & 449 & (17)  & 145 & (5)  & 64 & (3)\\[1ex]
  	& AIBT	& 56 & (3)    & 185 & (5)   & 87 & (3)   & 52 & (2)\\
  	& SS	& 98 & (3)    & 253 & (10)  & 99 & (4)   & 94 & (4)\\
  7	& CV	& 96 & (3)    & 441 & (25)  & 135 & (6)  & 54 & (3)\\
  	& BT	& 414 & (11)  & 1716 & (21) & 225 & (6)  & 57 & (2)\\
  	& FDR	& 294 & (18)  & 758 & (27)  & 253 & (9)  & 93 & (4)\\[1ex]
	& AIBT	& 535 & (21)  & 1023 & (15) & 448 & (18) & 153 & (6)\\
	& SS	& 482 & (13)  & 973 & (45)  & 399 & (14) & 147 & (3)\\
  3	& CV	& 452 & (11)  & 914 & (34)  & 375 & (13) & 148 & (6)\\
	& BT	& 860 & (24)  & 2015 & (37) & 448 & (12) & 140 & (4)\\
	& FDR	& 1230 & (52) & 2324 & (88) & 862 & (31) & 148 & (3)\\\hthickline
 \end{tabular}
\end{table}

The goodness of fit of each estimator was measured by its average
mean-square error (AMSE) over the 25 replications.  Table \ref{tab:sim}
presents the results.  It is clear that our estimator performs extremely
well with respect to the other estimators when the signal-to-noise ratio is
moderate or large, but less well, though still competitively, when there is a
small signal-to-noise ratio.  

\section{Conclusions and future work}\label{concl}

We have introduced a procedure for Bayesian wavelet thresholding which
uses the naturally clustered nature of the wavelet transform when deciding
how much weight to give coefficient values.  
In comparisons with other methods, our approach performed very
well for moderate and low noise levels, and reasonably competitively for
higher noise levels.

One possible area for future work would be to replace equation
\eqref{truecoef} with
\[
	d_{jk}|\xi \sim N(0,\tau^2(\xi_{jk})^z),
\]
where $z$ would be a further parameter.  This would modify the number of
points which are likely to be alive at any given location and thus also
modify the tail behaviour of the prior. 
The idea behind this suggestion is that when we know that the behaviour of
the data is either heavy or light tailed, we could adjust $z$ to compensate.
This could possibly also help speed up convergence by reducing the number
of points at locations with large values of $d_{jk}$.  As inclusion of this
extra parameter requires only minor modifications, the software
discussed actually includes this option.  The results presented in
Section~\ref{simstud} were generated by simply setting $z=1$.

A second possible area for future work would be to develop some automatic
methods for choosing the parameter values, perhaps using the method of
maximum pseudo-likelihood \citep{bes:spa-int,bes:sta-ana,bes:met-sta}.

Software implementing the work described in this paper is available on
request from the first author.

\section{Acknowledgements}

The first author would like to thank Guy Nason and Paul Northrop for helpful
discussions.

\bibliographystyle{hapalike}
\bibliography{exact,space,wave,books}

\end{document}